\documentclass[10pt,twoside]{article}
\usepackage{graphicx}
\usepackage{amsmath}
\usepackage{Latex-document}

\newtheorem{proposition}{Proposition}

\title{\bf \boldmath{$P \ne NP$}, Propositional Proof Complexity, \vskip -2mm and
Resolution Lower Bounds for the \vskip -2mm Weak Pigeonhole Principle \vskip 6mm}

\author{Ran Raz\vspace*{-0.5cm}\thanks{Department of Computer Science,
Weizmann Institute for Science, Rehovot 76100, Israel. E-mail:
ranraz@wisdom.weizmann.ac.il}}

\date{\vspace{-8mm}}

\begin{document}

\maketitle

\thispagestyle{first} \setcounter{page}{685}

\begin{abstract}

\vskip 3mm

Recent results established exponential lower bounds for the length
of any Resolution proof for the weak pigeonhole principle. More
formally, it was proved that any Resolution proof for the weak
pigeonhole principle, with $n$ holes and any number of pigeons, is
of length $\Omega(2^{n^{\epsilon}})$, (for a constant $\epsilon =
1/3$). One corollary is that certain propositional formulations of
the statement $P \ne NP$ do not have short Resolution proofs.
After a short introduction to the problem of $P \ne NP$ and to the
research area of propositional proof complexity, I will discuss
the above mentioned lower bounds for the weak pigeonhole principle
and the connections to the hardness of proving $P \ne NP$.

\vskip 4.5mm


\end{abstract}

\vskip 12mm

\section{Propositional logic} \label{section 1}\setzero
\vskip-5mm \hspace{5mm}

The basic syntactic units (atoms) of propositional logic are
Boolean variables $x_1,...,x_n \in \{0,1\}$, where the value 0
represents {\it False} and the value 1 represents {\it True}. The
propositional variables are combined with standard Boolean gates
(also called connectives), such as, {\it AND} (conjunction), {\it
OR} (disjunction), and {\it NOT} (negation), to form Boolean
formulas. Recall that in propositional logic there are no
quantifiers.

A {\em literal} is either an atom (i.e., a variable $x_i$) or the
negation of an atom (i.e., $\neg x_i$). A {\em clause} is a
disjunction of literals. A {\em term} is a conjunction of
literals. A formula $f$ is in {\it conjunctive-normal-form} (CNF)
if it is a conjunction of clauses. A  formula $f$ is in {\it
disjunctive-normal-form} (DNF) if it is a disjunction of terms.
Since there are standard ways to transform a formula to CNF or DNF
(by adding new variables), many times we limit the discussion to
CNF formulas or DNF formulas.

A Boolean formula $f(x_1,...,x_n)$ is a {\it tautology} if
$f(x_1,...,x_n)=1$ for every $x_1,...,x_n$. A Boolean formula
$f(x_1,...,x_n)$ is {\it unsatisfiable} if $f(x_1,...,x_n)=0$ for
every $x_1,...,x_n$. Obviously, $f$ is a tautology if and only if
$\neg f$ is unsatisfiable.

Given a formula $f(x_1,...,x_n)$, one can decide whether or not
$f$ is a tautology by checking all the possibilities for
assignments to $x_1,...,x_n$. However, the time needed for this
procedure is exponential in the number of variables, and hence may
be exponential in the length of the formula $f$.

\section{\boldmath{$P \ne NP$}} \label{section 2}\setzero
\vskip-5mm \hspace{5mm}

$P \ne NP$ is the central open problem in complexity theory and
one of the most important open problems in mathematics today. The
problem has thousands of  equivalent formulations. One of these
formulations is the following:
\begin{quote}
{\bf Is there a polynomial time algorithm $A$ that gets as input a
Boolean formula $f$ and outputs 1 if and only if $f$ is a
tautology~?}
\end{quote}
$P \ne NP$ states that there is no such algorithm.

A related open problem in complexity theory is the problem of $NP
\ne Co-NP$. The problem can be stated as follows:
\begin{quote}
{\bf Is there a polynomial time algorithm $A$ that gets as input a
Boolean formula $f$ and a string $z$, and such that: $f$ is a
tautology if and only if there exists $z$ s.t.:
\begin{enumerate}
\item The length of $z$ is at most polynomial in the length of $f$.
\item $A(f,z) =1$.
\end{enumerate}}
\end{quote}
$NP \ne Co-NP$ states that there is no such algorithm. Obviously,
$NP \ne Co-NP$ implies $P \ne NP$.

It is widely believed that $P \ne NP$ (and $NP \ne Co-NP$). At
this point, however, we are still far from giving a solution for
these problems. It is not clear why these problems are so hard to
solve.

\section{Propositional proof theory} \label{section 3}\setzero
\vskip-5mm \hspace{5mm}

Propositional proof theory is the study of the length of proofs
for different tautologies in different propositional proof
systems.

The notion of {\it propositional proof system} was introduced by
Cook and Reckhow in 1973, as a direction for proving $NP \ne
co-NP$ (and hence also $P \ne NP$)~\cite{CR}. A propositional
proof system is a polynomial time algorithm $A(f,z)$ such that a
Boolean formula $f$ is a tautology if and only if there exists $z$
such that $A(f,z) = 1$ (note that we do not require here that the
length of $z$ is at most polynomial in the length of $f$). We
think of the string $z$ as a proof for $f$ in the proof system
$A$. We say that a tautology $f$ is {\it hard} for a proof system
$A$ if any proof $z$ for $f$ in the proof system $A$ is of length
super-polynomial in the length of $f$.

Many times we prefer to talk about unsatisfiable formulas, rather
than tautologies, and about refutation systems, rather than proof
systems. A {\it propositional refutation system} is a polynomial
time algorithm $A(f,z)$ such that a Boolean formula $f$ is
unsatisfiable if and only if there exists $z$ such that $A(f,z) =
1$. We think of the string $z$ as a refutation for $f$ in the
refutation system $A$. We think of a refutation $z$ for $f$ also
as a proof for $\neg f$ (and vice versa).

It is easy to see that $NP \ne co-NP$ if and only if for every
propositional proof system $A$ there exists a hard tautology, that
is, a tautology $f$ with no short proofs. It was hence suggested
by Cook and Reckhow to study the length of proofs for different
tautologies in stronger and stronger propositional proof systems.
It turns out that in many cases these problems are very
interesting in their own right and are related to many other
interesting problems in complexity theory and in logic, in
particular when the tautology $f$ represents a fundamental
mathematical principle.

For a recent survey on the main research directions in
propositional proof theory, see~\cite{BP2}.

\section{Resolution} \label{section 4}\setzero
\vskip-5mm \hspace{5mm}

{\em Resolution} is one of the simplest and most widely studied
propositional proof systems. Besides its mathematical simplicity
and elegance, Resolution is a very interesting proof system also
because it generalizes the Davis-Putnam procedure and several
other well known proof-search procedures. Moreover, Resolution is
the base for most automat theorem provers existing today.

The {\em Resolution rule} says that if $C$ and $D$ are two clauses
and $x_i$ is a variable then any assignment (to the variables
$x_1,...,x_n$) that satisfies both of the clauses, $C \lor x_i$
and $D \lor \neg x_i$, also satisfies the clause $C \vee D$. The
clause $C \vee D$ is called the {\em resolvent} of the clauses $C
\vee x_i$  and $D \vee \neg x_i$  on the variable $x_i$.

Resolution is usually presented as a propositional refutation
system for CNF formulas. Since there are standard ways to
transform a formula to CNF (by adding new variables), this
presentation is general enough. A {\em Resolution refutation} for
a CNF formula $f$ is a sequence of clauses  $C_1,C_2,\ldots,C_s$,
such that:
\begin{enumerate}
\item Each clause $C_j$ is either a clause of $f$ or a
resolvent of two previous clauses in the sequence.
\item The last
clause, $C_s$, is the empty clause.
\end{enumerate}
We think of the empty clause as a clause that has no satisfying
assignments, and hence a contradiction was obtained.

We think of a Resolution refutation for $f$ also as a proof for
$\neg f$. Without loss of generality, we assume that no clause in
a Resolution proof contains both $x_i$ and $\neg x_i$ (such a
clause is always satisfied and hence it can be removed from the
proof). The {\em length}, or {\em size}, of a Resolution proof is
the number of clauses in it.

We can represent a Resolution proof as an acyclic directed graph
on vertices $C_1,\ldots,C_s$, where each clause of $f$ has
out-degree 0, and any other clause has two edges pointing to the
two clauses that were used to produce it.

It is well known that Resolution is a refutation system. That is,
a CNF formula $f$ is unsatisfiable if and only if there exists a
Resolution refutation for $f$. A well-known and widely studied
restricted version of Resolution (that is still a refutation
system) is called {\em Regular Resolution}. In a Regular
Resolution refutation, along any path in the directed acyclic
graph, each variable is resolved upon at most once.

\section{Resolution as a search problem} \label{section 5}\setzero
\vskip-5mm \hspace{5mm}

As mentioned above, we represent a Resolution proof as an acyclic
directed graph $G$ on the vertices $C_1,\ldots,C_s$. In this
graph, each clause $C_j$ which is an original clause of $f$ has
out-degree 0, and any other clause has two edges pointing to the
two clauses that were used to produce it. We call the vertices of
out-degree 0 (i.e., the clauses that are original clauses of $f$)
the {\em leaves} of the graph. Without loss of generality, we can
assume that the only clause with in-degree 0 is the last clause
$C_s$ (as we can just remove any other clause with in-degree 0).
We call the vertex $C_s$ the {\em root} of the graph.

We label each vertex $C_j$ in the graph by the variable $x_i$ that
was used to derive it (i.e., the variable $x_i$ that was resolved
upon), unless the clause $C_j$ is an original clause of $f$ (and
then $C_j$ is not labelled). If a clause $C_j$ is labelled by a
variable $x_i$ we label the two edges going out from $C_j$ by 0
and 1, where the edge pointing to the clause that contains $x_i$
is labelled by 0, and the edge pointing to the clause that
contains $\neg x_i$ is labelled by 1. That is, if the clause $C
\vee D$ was derived from the two clauses $C \lor x_i$ and $D \lor
\neg x_i$ then the vertex $C \vee D$ is labelled by $x_i$, the
edge from the vertex $C \vee D$ to the vertex $C \lor x_i$ is
labelled by 0 and the edge from the vertex $C \vee D$ to the
vertex $D \lor \neg x_i$ is labelled by 1. For a non-leaf node $u$
of the graph $G$, define,
\begin{quote}
${\bf Label(u)} =$ the variable
labelling $u$.
\end{quote}
We think of $Label(u)$ as a variable queried at the node $u$.

Let $p$ be a path on $G$, starting from the root. Note that along
a path $p$, a variable $x_i$ may appear (as a label of a node $u$)
more than once. We say that the path $p$ evaluates $x_i$ to 0 if
$x_i=Label(u)$ for some node $u$ on the path $p$, and after the
last appearance of $x_i$ as $Label(u)$ (of a node $u$ on the path)
the path $p$ continues on the edge labelled by 0 (i.e., if $u$ is
the last node on $p$ such that $x_i=Label(u)$ then $p$ contains
the edge labelled by 0 that goes out from $u$). In the same way,
we say that the path $p$ evaluates $x_i$ to 1 if $x_i=Label(u)$
for some node $u$ on the path $p$, and after the last appearance
of $x_i$ as $Label(u)$ (of a node $u$ on the path) the path $p$
continues on the edge labelled by 1 (i.e., if $u$ is the last node
on $p$ such that $x_i=Label(u)$ then $p$ contains the edge
labelled by 1 that goes out from $u$).

For any node $u$ of the graph $G$, we define $Zeros(u)$ to be the
set of variables that the node $u$ ``remembers'' to be 0, and
$Ones(u)$ to be the set of variables that the node $u$
``remembers'' to be 1, that is,
\begin{quote}
${\bf Zeros(u)} =$ the set of variables
that are evaluated to 0 by every path $p$ from the root to $u$.
\end{quote}
\begin{quote}
${\bf Ones(u)} =$ the set of variables
that are evaluated to 1 by every path $p$ from the root to $u$.
\end{quote}
Note that for any $u$, the two sets $Zeros(u)$ and $Ones(u)$ are
disjoint.

The following proposition gives the connection between the sets
$Zeros(u)$, $Ones(u)$ and the literals appearing in the clause
$u$. The proposition is particularly interesting when $u$ is a
leaf of the graph.

\begin{proposition} \label{pro:search}
Let $f$ be an unsatisfiable CNF formula and let $G$ be (the graph
representation of) a Resolution refutation for $f$. Then, for any
node $u$ of $G$ and for any $x_i$, if the literal $x_i$ appears in
the clause $u$ then $x_i \in Zeros(u)$, and if the literal $\neg
x_i$ appears in the clause $u$ then $x_i \in Ones(u)$.
\end{proposition}

\section{The weak pigeonhole principle} \label{section 6}\setzero
\vskip-5mm \hspace{5mm}

The {\em Pigeonhole Principle} (PHP) is probably the most widely
studied tautology in propositional proof theory. The tautology
$PHP_n$ is a DNF encoding of the following statement: There is no
one to one mapping from $n+1$ pigeons to $n$ holes. The {\em Weak
Pigeonhole Principle} (WPHP) is a version of the pigeonhole
principle that allows a larger number of pigeons. The tautology
$WPHP^m_n$ (for $m \geq n+1$) is a DNF encoding of the following
statement: There is no one to one mapping from $m$ pigeons to $n$
holes. For $m>n+1$, the weak pigeonhole principle is a weaker
statement than the pigeonhole principle. Hence, it may have much
shorter proofs in certain proof systems.

The weak pigeonhole principle is one of the most fundamental
combinatorial principles. In particular, it is used in most
probabilistic counting arguments and hence in many combinatorial
proofs. Moreover, as observed by Razborov, there are certain
connections between the weak pigeonhole principle and the problem
of $P \ne NP$ \cite{Razb3}. Indeed, the weak pigeonhole principle
(with a relatively large number of pigeons) can be interpreted as
a certain encoding of the following statement: There are no small
DNF formulas for $SAT$ (where $SAT$ is the satisfiability
problem). Hence, in most proof systems, a short proof for certain
formulations of the statement ``There are no small formulas for
$SAT$'' can be translated into a short proof for the weak
pigeonhole principle. That is, a lower bound for the length of
proofs for the weak pigeonhole principle usually implies a lower
bound for the length of proofs for certain formulations of the
statement $P \ne NP$. While this doesn't say much about the
problem of $P \ne NP$, it does demonstrate the applicability and
relevance of the weak pigeonhole principle for other interesting
problems.

Formally, the formula $WPHP^m_n$ is expressed in the following
way. The underlying Boolean variables, $x_{i,j}$, for $1 \leq i
\leq m$ and $1 \leq j \leq n$, represent whether or not pigeon $i$
is mapped to hole $j$. The negation of the pigeonhole principle,
$\neg WPHP^m_n$, is expressed as the conjunction of $m$ {\em
pigeon clauses} and ${m \choose 2} \cdot n$ {\em hole clauses}.
For every $1 \leq i \leq m$, we have a pigeon clause, $$(x_{i,1}
\lor \ldots \lor x_{i,n}),$$ stating that pigeon $i$ maps to some
hole. For every $1 \leq i_1 < i_2 \leq m$ and every $1 \leq j \leq
n$, we have a hole clause,
$$(\neg x_{i_1,j} \lor \neg x_{i_2,j}),$$
stating that pigeons $i_1$ and $i_2$ do not both map to hole $j$.
We refer to the pigeon clauses and the hole clauses also as pigeon
axioms and hole axioms. Note that $\neg WPHP^m_n$ is a CNF
formula.

Let $G$ be (the graph representation of) a Resolution refutation
for $\neg WPHP^m_n$. Then, by Proposition~\ref{pro:search}, for
any leaf $u$ of the graph $G$, one of the following is satisfied:
\begin{enumerate}
\item
$u$ is a pigeon axiom, and then for some $1 \leq i \leq m$, the
variables $x_{i,1}, \ldots , x_{i,n}$ are all contained in
$Zeros(u)$.
\item
$u$ is a hole axiom, and then for some $1 \leq j \leq n$, there
exist two different variables $x_{i_1,j},x_{i_2,j}$ in $Ones(u)$.
\end{enumerate}

\section{Resolution lower bounds for the weak pigeonhole principle}
\label{section 7}\setzero \vskip-5mm \hspace{5mm}

There are trivial Resolution proofs (and Regular Resolution
proofs) of length $2^{n} \cdot poly(n)$ for the pigeonhole
principle and for the weak pigeonhole principle. In a seminal
paper, Haken proved that for the pigeonhole principle, the trivial
proof is (almost) the best possible \cite{H}. More specifically,
Haken proved that any Resolution proof for the tautology $PHP_n$
is of length $2^{\Omega(n)}$. Haken's argument was further
developed in several other papers (e.g., \cite{Urq,BP,BSW}). In
particular, it was shown that a similar argument gives lower
bounds also for the weak pigeonhole principle, but only for small
values of $m$. More specifically, super-polynomial lower bounds
were proved for any Resolution proof for the tautology $WPHP^m_n$,
for $m  < c \cdot n^2/ \log n$ (for some constant $c$) \cite{BT}.

For the weak pigeonhole principle with large values of $m$, there
do exist Resolution proofs (and Regular Resolution proofs) which
are much shorter than the trivial ones. In particular, it was
proved by Buss and Pitassi that for $m > c^{\sqrt{n \log n}}$ (for
some constant $c$), there are Resolution (and Regular Resolution)
proofs of length $poly(m)$ for the tautology $WPHP^m_n$
\cite{BussP}. Can this upper bound be further improved~? Can one
prove a matching lower bound~? A partial progress was made by
Razborov, Wigderson and Yao, who proved exponential lower bounds
for Regular Resolution proofs, but only when the Regular
Resolution proof is of a certain restricted form \cite{RWY}.

The weak pigeonhole principle with large number of pigeons has
attracted a lot of attention in recent years. However, the
standard techniques for proving lower bounds for Resolution failed
to give lower bounds for the weak pigeonhole principle. In
particular, for $m \geq n^2$, no non-trivial lower bound was known
until very recently.

In the last two years, these problems were completely solved. An
exponential lower bound for any Regular Resolution proof was
proved in~\cite{PR}, and an exponential lower bound for any
Resolution proof was finally proved in~\cite{Raz}. More precisely,
it was proved in~\cite{Raz} that for any $m$, any Resolution proof
for the weak pigeonhole principle $WPHP^m_n$ is of length
$\Omega(2^{n^{\epsilon}})$, where $\epsilon > 0$ is some global
constant ($\epsilon \approx 1/8$).

The lower bound was further improved in several results by
Razborov. The first result~\cite{Razb4} presents a proof for an
improved lower bound of $\Omega(2^{n^{\epsilon}})$, for $\epsilon
= 1/3$. The second result~\cite{Razb5} extends the lower bound to
an important variant of the pigeonhole principle, the so called
{\it weak functional pigeonhole principle}, where we require in
addition that each pigeon goes to exactly one hole. The third
result~\cite{Razb6} extends the lower bound to another important
variant of the pigeonhole principle, the so called {\it weak
functional onto pigeonhole principle}, where we require in
addition that every hole is occupied.

For a recent survey on the propositional proof complexity of the
pigeonhole principle, see~\cite{Razb7}.

\section{Lower bounds for \boldmath{$P \ne NP$}}
\label{section 8}\setzero \vskip-5mm \hspace{5mm}

Propositional versions of the statement $P \ne NP$ were introduced
by Razborov in 1995~\cite{Razb1} (see also~\cite{Razb2}). Razborov
suggested to try to prove super-polynomial lower bounds for the
length of proofs for these statements in stronger and stronger
propositional proof systems. This was suggested as a step for
proving the hardness of proving $P \ne NP$. The above mentioned
results for the weak pigeonhole principle establish such
super-polynomial lower bounds for Resolution.

Let $g: \{0,1\}^d \rightarrow \{0,1\}$ be a Boolean function. For
example, we can take $g=SAT$, where $SAT: \{0,1\}^d \rightarrow
\{0,1\}$ is the satisfiability function (or we can take any other
$NP$-hard function). We assume that we are given the truth table
of $g$. Let $t \leq 2^d$ be some integer. We think of $t$ as a
large polynomial in $d$, say $t=d^{1000}$.

Razborov suggested to study propositional formulations of the
following statement (in the variables $\vec{Z}$):
\begin{quote}
{\bf $\vec{Z}$ is (an encoding of) a Boolean circuit of size $t$
$\Longrightarrow$ \\
$\vec{Z}$ does not compute the function $g$.}
\end{quote}
Note that since the truth table of $g$ is of length $2^d$, a
propositional formulation of this statement will be of length at
least $2^{d}$, and it is not hard to see that there are ways to
write this statement as a DNF formula of length $2^{O(d)}$ (and
hence, its negation is a CNF formula of that length). The standard
way to do that is by including in $\vec{Z}$ both, the
(topological) description of the Boolean circuit, as well as the
value that each gate in the circuit outputs on each input for the
circuit.

In~\cite{Razb3}, Razborov presented a lower bound for the degree
of {\it Polynomial Calculus} proofs for the weak pigeonhole
principle, and used this result to prove a lower bound for the
degree of Polynomial Calculus proofs for a certain version of the
above statement. Following this line of research, it was proved
in~\cite{Raz,Razb6} (in a similar way) that if $t$ is a large
enough polynomial in $d$ (say $t=d^{1000}$) then any Resolution
proof for certain versions of the above statement is of length
super-polynomial in $2^d$, that is, super-polynomial in the length
of the statement.

In particular, this can be interpreted as a super-polynomial lower
bound for Resolution proofs for certain formulations of the
statement $P \ne NP$ (or, more precisely, of the statement $NP
\not \subset P/poly$).

It turns out that the exact way to give the (topological)
description of the circuit is also important in some cases. This
was done slightly differently in~\cite{Raz} and in~\cite{Razb6}.
In~\cite{Raz}, $\vec{Z}$ was used to encode a Boolean circuit of
unbounded fan-in, whereas~\cite{Razb6} considered Boolean circuits
of fan-in 2. It turns out that for the stronger case of unbounded
fan-in, the lower bound for the weak pigeonhole principle is
enough~\cite{Raz}, whereas for the weaker case of fan-in 2 one
needs the lower bound for the weak functional onto pigeonhole
principle~\cite{Razb6} (in fact, this was one of the main
motivations to consider the onto functional case). Otherwise, the
proof seems to be quite robust in the way the Boolean circuit is
encoded.

\noindent{\bf Acknowledgement.}\  I would like to thank Toni
Pitassi for very enjoying collaboration that lead to the results
in~\cite{PR,Raz}.

\label{lastpage}


\begin{thebibliography}{aa}

\bibitem{BP} Beame, P., and Pitassi, T.,
``Simplified and improved resolution lower bounds,'' {\it
Foundations of Computer Science}, 1996, 274--282.

\bibitem{BP2} Beame, P., and Pitassi, T.,
``Propositional Proof Complexity: Past, Present, and Future,''
{\it Current Trends in Theoretical Computer Science}, 2001,
42--70.

\bibitem{BussP} Buss, S., and Pitassi, T.,
``Resolution and the weak pigeonhole principle,'' {\it Lecture
Notes in Computer Science, Springer-Verlag}, vol. 1414, 1998,
149--156. (Selected Papers of Computer Science Logic 11th
International Workshop, 1997).

\bibitem{BSW} Ben-Sasson, E., and Wigderson, A.,
``Short proofs are narrow--resolution made simple,'' {\it Journal
of the ACM}, 48(2),2001, 149--168.

\bibitem{BT} Buss, S., and Turan, G.,
``Resolution proofs of generalized pigeonhole principles,'' {\it
Theoretical Computer Science}, 62(3), 1988, 311--317.

\bibitem{CR} Cook, S., and Reckhow R.,
``The relative efficiency of propositional proof systems,'' {\it
Journal of Symbolic Logic}, 44(1), 1979, 36--50.

\bibitem{H} Haken, A.,
``The intractability of resolution,'' {\it Theoretical Computer
Science}, 39(2-3), 1985, 297--308.

\bibitem{PR} Pitassi, T., and Raz, R.,
``Regular resolution lower bounds for the weak pigeonhole
principle,'' {\it Symposium on Theory of Computing}, 2001,
347--355.

\bibitem{Raz} Raz, R.,
``Resolution lower bounds for the weak pigeonhole principle,''
{\it Symposium on Theory of Computing}, 2002.

\bibitem{Razb1} Razborov, A.,
``Bounded arithmetic and lower bounds in Boolean complexity,''
{\it Feasible Mathematics II. Progress in Computer Science and
Applied Logic}, vol. 13, 1995, 344--386.

\bibitem{Razb2} Razborov, A.,
``Lower bounds for propositional proofs and independence results
in Bounded Arithmetic,'' {\it Lecture Notes in Computer Science,
Springer Verlag}, vol. 1099, 1996, 48--62. (Proc. of the 23rd
ICALP).

\bibitem{Razb3} Razborov, A.,
``Lower bounds for the polynomial calculus,'' {\it Computational
Complexity}, 7(4), 1998, 291--324.

\bibitem{Razb4} Razborov, A.,
``Improved resolution lower bounds for the weak pigeonhole
principle,'' {\it Electronic Colloquium on Computational
Complexity (ECCC)}, 8(055), 2001.

\bibitem{Razb5} Razborov, A.,
``Resolution lower bounds for the weak functional pigeonhole
principle,'' {\it Electronic Colloquium on Computational
Complexity (ECCC)}, 8(075), 2001. (to appear in {\it Theoretical
Computer Science}).

\bibitem{Razb6} Razborov, A.,
``Resolution Lower Bounds for Perfect Matching Principles,'' {\it
Proc. of the 17th IEEE Conference on Computational Complexity},
2002.

\bibitem{Razb7} Razborov, A.,
``Proof Complexity of Pigeonhole Principles,'' {\it Developments
in Language Theory}, 2001, 100--116.

\bibitem{RWY} Razborov, A., Wigderson, A., and Yao, A.,
``Read-once branching programs, rectangular proofs of the
pigeonhole principle, and the transversal calculus,'' {\it
Symposium on Theory of Computing}, 1997, 739--748.

\bibitem{Urq} Urquhart, A.,
``Hard examples for resolution,'' {\it Journal of the ACM}, vol.
34, 1987, 209--219.

\end{thebibliography}
\end{document}